\documentclass[pdflatex,sn-basic]{sn-jnl}

\usepackage{graphicx}%
\usepackage{multirow}%
\usepackage{amsmath,amssymb,amsfonts}%
\usepackage{amsthm}%
\usepackage{mathrsfs}%
\usepackage[title]{appendix}%
\usepackage{xcolor}%
\usepackage{textcomp}%
\usepackage{manyfoot}%
\usepackage{booktabs}%
\usepackage{algorithm}%
\usepackage{algorithmicx}%
\usepackage{algpseudocode}%
\usepackage{listings}%
\usepackage{url}%

\theoremstyle{thmstyleone}%
\theoremstyle{thmstyletwo}%

\theoremstyle{thmstylethree}%

\begin{document}

\title[The Fabricated Front]{The Fabricated Front: Generative AI and the Opacity of Workplace Performance}

\author*[1]{\fnm{Tom} \sur{Van Nuenen}}\email{tomvannuenen@berkeley.edu}

\author[1,2]{\fnm{Pratik} \sur{S. Sachdeva}}\email{pratik.sachdeva@berkeley.edu}

\author[1]{\fnm{Sahiba} \sur{Chopra}}\email{sahiba.chopra@berkeley.edu}

\affil*[1]{\orgdiv{D-Lab}, \orgname{University of California, Berkeley}, 
\orgaddress{\street{356 Social Science Building}, \city{Berkeley}, 
\postcode{94720}, \state{CA}, \country{USA}}}

\affil[2]{\orgdiv{Haas School of Business}, \orgname{University of California, Berkeley}, 
\orgaddress{\street{2220 Piedmont Avenue}, \city{Berkeley}, 
\postcode{94720}, \state{CA}, \country{USA}}}


\abstract{Generative AI (GenAI) has become a fixture of workplace life. Current research asks chiefly what this implies for jobs and outputs, measured in productivity, displacement, or bias. What remains underexamined are the interactional reconfigurations that GenAI produces at work. The emerging concept of effort opacity has begun to fill this gap by highlighting the systematic decoupling of observable output from human engagement. When GenAI makes interactional cues less diagnostic, it weakens the reciprocal exchange that sustains collaborative trust. Extending this account of effort opacity, we examine the interactional mechanics that produce opacity in everyday workplace encounters. 

Drawing on Erving Goffman's dramaturgical framework and 1,250 interview transcripts from Anthropic's AI Interviewer dataset, we identify five opacity mechanisms through which workplace fronts are reorganized: \textit{voice} (whose stance the words index), \textit{provenance} (who can stand behind the artifact), \textit{vulnerability} (whether the worker is uncertain), \textit{attention} (whether the worker is engaged), and \textit{investment} (how much labor the output reflects). We show that professionals defend the identity mechanisms while freely producing opacity around the labor mechanisms, and trace this asymmetry to the output-centered organization of contemporary work, where deliverables already stand in for the labor process that produced them. The governance task, accordingly, is one of involvement management: specifying which forms of human involvement (attention, effort, judgment) must remain inspectable, and to whom. Workplace AI policies built on universal disclosure will systematically misrecognize a social field in which inspectability is already audience-relative.}

\keywords{Generative AI, Workplace communication, Effort opacity, Interactional opacity, Workplace trust, AI-mediated communication, Professional identity}



\maketitle

\section{Introduction}

Workplace trust depends on the everyday interpretation of communicative performances. When a colleague sends an email or delivers a report, we make inferences about the person behind the performance. Word choice signals attentiveness, for instance, and distinctive phrasing indexes personality. These diagnostic cues form the tacit infrastructure of workplace coordination, helping us allocate trust and calibrate expectations. In other words, professional communication functions as a ``front,'' a patterned performance through which workers make competence and commitment available for interpretation \citep{goffman1959presentation}. Generative AI (GenAI) disrupts this infrastructure by introducing a new form of mediated communication in which a system can modify, augment, or generate interpersonal messages on behalf of a communicator \citep{hancock2020aimediated}. When outputs can be produced without the engagement they appear to represent, the interpretive framework through which we read professional contributions becomes unreliable.

This paper examines how professionals navigate the use of an AI-mediated front. We take as our starting point the concept of ``effort opacity,'' the systematic decoupling of observable output from human engagement \citep{chopra2026engagement}. The concept condenses a longer lineage in organizational scholarship, which has treated much of work as invisible by design \citep{star1999layers}, output as a signal of the diligence behind it \citep{feldman1981information}, and visibility as a managed, consequential accomplishment \citep{bernstein2012transparency, leonardi2020behavioral}. When GenAI makes interactional cues less diagnostic, it weakens the reciprocal exchange that sustains collaborative trust. A remaining question is how professionals decide which forms of legibility to preserve and how those choices reorganize accountability in everyday work.

Prior work has shown that AI use can carry reputational costs and create a transparency dilemma, in which disclosure may reduce rather than restore trust \citep{schilke2025transparency, reif2025social}. We extend the reputational-cost and disclosure-dilemma literatures through a typology of opacity mechanisms, each describing an inference audiences ordinarily draw from a worker's front that AI use now disrupts.
Drawing on Erving Goffman's dramaturgical framework, we argue that GenAI fundamentally alters the shared assumptions about who is speaking in what capacity, what commitments utterances carry, and what counts as evidence of engagement \citep{goffman1959presentation}. 

We analyze 1,250 interview transcripts from Anthropic's AI Interviewer dataset, capturing professionals across the general workforce, the creative industry, and the sciences. We show how these workers resist opacity around identity mechanisms (who one appears to be), yet freely produce it around labor mechanisms (what one actually did). We argue that AI policies that ignore the audience structure of inspectability will produce compliant disclosures, while the concealment they aim to address will continue.

\section{From Effort Opacity to Interactional Opacity}

Erving Goffman's dramaturgical sociology treats social life as performances in which actors control the impressions they convey \citep{goffman1959presentation}. Central is the ``definition of the situation'': an understanding, shared by participants, about what kind of interaction is taking place, who is participating in what capacity, and what standards of conduct therefore apply \citep{goffman1974frame}. In workplace settings, this includes assumptions about whether an email, report, or comment reflects a person's own judgment, how much thought or attention it represents, and what kind of accountability attaches to it.

To specify how GenAI unsettles this definition of the situation, we draw on three related Goffmanian concepts. First, \textit{footing} describes speakers' alignment toward utterances \citep{goffman1981forms}. Goffman distinguishes the \textit{animator} (who produces an utterance), the \textit{author} (who composes it), and the \textit{principal} (whose position it establishes). These roles, Goffman noted, are routinely separable in institutional contexts: a press secretary, for instance, animates words composed by speechwriters on behalf of the office they represent. Such signals, like a podium with the presidential seal, mark the decoupling. Generative AI inserts the same animator/author split into ordinary contexts where no such signals exist and author and animator are assumed identical. 

Second, \textit{front stage} and \textit{back stage} capture how performers manage information across regions visible and invisible to the audience. Goffman's example of a back stage is a collective space where teammates coordinate and rehearse for the front stage performance. AI enters this region as a substitute for the teammates who would have otherwise been there, and as a shared resource among peers. In both cases, GenAI reshapes what front stage performances seem to reveal about the labor behind them. Third, \textit{stigma} examines how individuals manage potentially discrediting attributes \citep{goffman1963stigma}. Emerging evidence suggests AI use is becoming such an attribute in some professional settings, resulting in strategic concealment \citep{schilke2025transparency}.

Existing work shows that GenAI increases productivity across writing, knowledge-work, and service tasks, with the largest gains for less-experienced and lower-performing workers \citep{noy2023experimental,dellacqua2023navigating,brynjolfsson2025generative}. Yet these are also the workers whose unaided competence is most under scrutiny, and who therefore have the most reason to conceal AI use. Disclosure can invite judgments of lower competence or motivation \citep{reif2025social} and may reduce rather than restore trust \citep{schilke2025transparency}. This matches a long line of work showing that audiences read visible output and displayed effort as signals of competence: conspicuous information use symbolizes competent decision-making \citep{feldman1981information}, and busyness itself confers status \citep{bellezza2017conspicuous}. In the GenAI workplace this management is already routine: workers scrub the cues of AI use both to avoid stigma and because seamless, undetected use itself signals expertise \citep{xia2026clever}. Underlying these dynamics is a broader shift. As AI makes outputs easier to produce, observable performance becomes less diagnostic of the engagement behind it \citep{chopra2026engagement}.

These studies, however, register the decoupling mainly as a matter of individual costs and evaluations. What it does to the interactional order of the workplace, the shared assumptions through which colleagues read one another's performances, is less well-known. The account closest to ours, \citet{klowait2025presentation} treats LLMs as supplementary rather than substitutive automation: machines absorb routine cognitive tasks while humans preserve a visible relational performance. We extend this account empirically: Goffman's framework lets us name three features of ordinary workplace performance, and our close readings in \S 4 show each being undercut by GenAI. The social occasions Goffman describes, such as the moments when suspicion leads to repair or novice uncertainty becomes an opportunity for instruction, dissolve before they can happen. Footing splits across time, with workers using GenAI to animate, in the present, what an archived version of themselves would have written. And what Goffman treats as a continuous stream of small choices collapses into a single founding decision. We will argue these breaks reframe the analytic problem from impression management to what we will call \textit{involvement management}: the practical question of which forms of human involvement must remain inspectable for the performance to count as the worker's own.

A key implication running throughout the analysis is that the opacity of labor should not be mistaken for its disappearance. In many cases, the labor AI removes from the visible artifact reappears as invisible upkeep---prompting, checking, hedging, and repairing the cue surface---redirecting effort from producing the output to maintaining the conditions under which it still reads as the worker's own. This is invisible work in the classic sense: labor an organization depends on but does not render visible or creditable \citep{star1999layers}.

\section{Five Opacity Mechanisms}

Building on Chopra's effort opacity, i.e. the general decoupling of output from human engagement, we describe five mechanisms that specify where the decoupling occurs in interaction. 
\textit{Voice opacity} concerns the filtering of stance cues and footing: AI-mediated text may be eloquent but no longer indexes the person who sent it, removing both the personality cues through which colleagues recognize one another \citep{hyland2012disciplinary} and the situational cues (mood, urgency, hesitation) that signal how they are approaching the encounter. \textit{Vulnerability opacity} emerges when AI handles uncertainty for workers, filtering out the hedges and pauses that ordinarily invite mentorship \citep{edmondson1999psychological}. \textit{Provenance opacity} describes gaps in accountability when workers cannot reconstruct the reasoning behind AI-assisted artifacts. For instance, direct AI content generation weakens writers' felt accountability for the resulting text \citep{li2024writing}. A work report, in our case, may appear defensible until others demand explanation or repair. \textit{Attention opacity} decouples interactional participation from cognitive engagement, with AI summarizers and meeting assistants letting workers perform attendance without attention while reshaping the official memory of events \citep{suchman1987plans}. Finally, \textit{Investment opacity} describes the process of high- and low-investment outputs becoming indistinguishable, shifting the basis for credit from demonstrated work to inferences about character and motive.

The five mechanisms are not parallel choices. Voice and provenance index who one appears to be: the personality and authorial stance audiences read from a worker's cue surface. Vulnerability, attention, and investment index what one actually did: the engagement, attentiveness, and effort the worker brought to the artifact. The typology specifies what existing accounts treat as undifferentiated. These mechanisms differ in terms of which audiences can detect them, in what kinds of evidence they leave behind, and in how workers strategically prioritize them. 

\section{Evidence from the Interview Corpus}

We analyze 1,250 transcripts from Anthropic's AI Interviewer dataset \citep{anthropic2025interviewer}, which includes interviews with general workforce participants ($n=1{,}000$), creative professionals ($n=125$), and scientists ($n=125$). Participants were recruited through crowdworker platforms; the workforce sample is led by education and instruction (17\%), computing and mathematics (16\%), and arts, design, and media (14\%), the creative sample by writers (48\%) and visual artists (21\%), and the scientist sample spans more than fifty disciplines. Transcripts are public under CC-BY with participants' consent. Each interview is a 10--15 minute AI-conducted conversation about GenAI use at work, following a question plan reviewed and finalized by Anthropic researchers before deployment. Using LLM-assisted coding (GPT-5.4, ``medium'' reasoning effort), we classified each transcript by opacity mechanism, salience (none, potential, clear), and behavioral orientation (production, avoidance, mixed). Salience captures whether a mechanism is absent, possible, or explicitly named; orientation captures whether the participant resists opacity, produces it, or does both. 

We hand-coded a stratified sample of 207 cases for accountability- and efficiency-oriented justifications. Group differences were tested with Pearson chi-squared tests and Bonferroni-adjusted post-hoc comparisons. Pairwise post-hoc comparisons used chi-squared tests with Yates' continuity correction and Bonferroni adjustment across three comparisons. We report Cram\'er's $V$ alongside each chi-squared statistic. Group comparisons treat each transcript as a single observation. For the identity/labor comparison in \S 4.2, which pools mechanism-level observations across transcripts, we account for within-transcript dependence using two cluster-robust checks: a GEE logistic regression with an exchangeable working correlation and transcript-clustered standard errors, and a transcript-level cluster bootstrap. 

Salience indicates how directly a mechanism appears in a transcript. We code this as \textit{absent} when the participant does not describe anything related to it; \textit{potential} when the participant describes an activity that could surface as opacity (given the right interactional conditions) but does not explicitly name it as such; and \textit{clear} when the participant explicitly mentions the opacity concern.
 
Behavioral orientation captures the participant's stance toward the mechanism. \textit{Avoidance} describes practices that resist producing opacity, such as a worker editing AI output to sound like themselves before sending. \textit{Production} describes practices that create or allow opacity, such as a worker delegating the cue surface to AI without correction. Mixed cases include both within the same account. A worker who consistently rewrites AI-drafted emails to preserve their style is coded as \textit{voice opacity}, \textit{clear}, \textit{avoidance}. A worker who sends AI-generated text without editing codes as \textit{clear}, \textit{production}. A worker who uses AI for emails without addressing whether the resulting text reflects their voice is coded as \textit{potential}.

To assess the reliability of the LLM-assisted coding, two coders hand-coded the same 50 transcripts from a stratified random subsample (40 workforce, 5 creative, 5 science) on all five mechanisms, blind to the model's labels. Each made an independent first pass, after which a subset of cases was discussed and the codes were finalized. The finalized codes are highly consistent between coders (pooled salience $\kappa = 0.94$; orientation $\kappa = 1.00$ among jointly present cases), though because some disagreements were discussed, we treat this as consistency of the finalized codes rather than independent inter-rater reliability. Against the model's labels, agreement was substantial for both coders: 83.2\% and 85.6\% of the 250 salience judgments agreed exactly (Cohen's $\kappa = 0.66$ and $0.70$; linear-weighted $\kappa = 0.71$ and $0.75$; binary presence $\kappa = 0.72$ and $0.74$). 95--100\% of disagreements were a single step apart on the ordinal scale, and were concentrated in provenance, where both coders apply a somewhat more inclusive threshold than the model ($\kappa = 0.24$ and $0.26$). Agreement with the model on behavioral orientation among jointly present cases was substantial ($\kappa = 0.66$ and $0.68$). The identity/labor asymmetry discussed in \S 4.2 replicates in each coder's codes (OR $= 3.8$, Fisher exact $p = .034$, and OR $= 7.0$, $p = .010$; the model's codes on the same transcripts give OR $= 6.8$, $p = .006$).

\subsection{Mechanism Prevalence and Variation Across Groups}

\begin{figure}[htbp]
\centering
\includegraphics[width=0.9\textwidth]{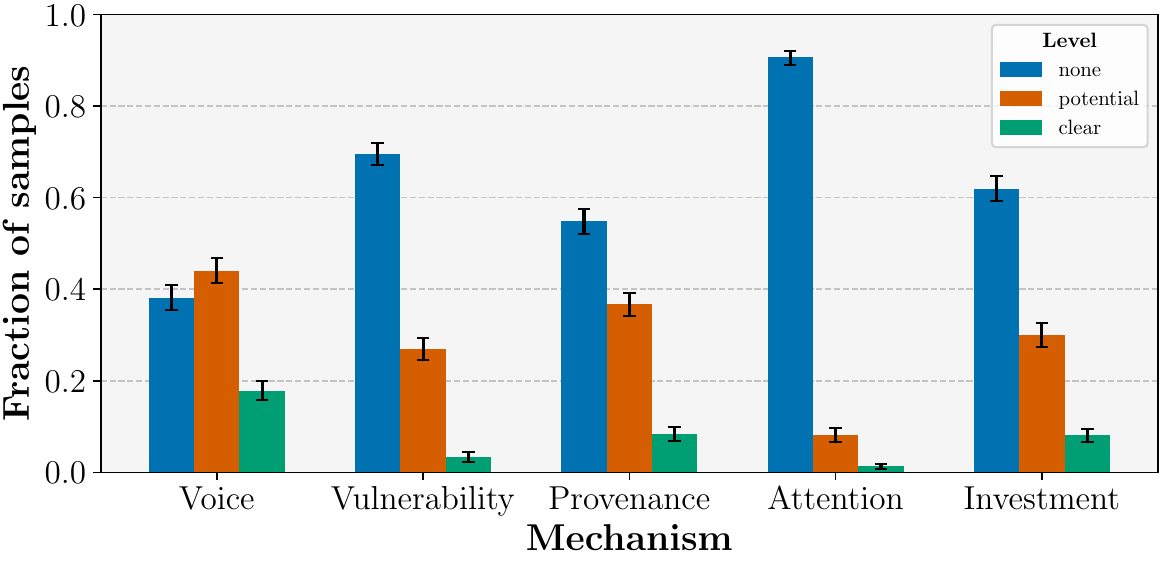}
\caption{Prevalence of opacity mechanisms across all interviews ($N = 1,250$). Bars show fraction of transcripts at each level (none, potential, clear).}
\label{fig:prevalence}
\end{figure}

Our coding reveals that opacity concerns pervade professionals' accounts but vary substantially in salience (Figure~\ref{fig:prevalence}). Voice opacity is most prevalent (61.8\% of transcripts), followed by provenance (45.1\%), investment (38.1\%), vulnerability (30.4\%), and attention (9.4\%). Attention's rarity reflects both the recency of AI meeting tools and the narrower range of tasks where engagement can be meaningfully delegated. Where it does appear, accounts frame delegation as efficiency rather than evasion, contrasting with the moral vocabulary that surrounds other mechanisms.

\begin{figure}[htbp]
\centering
\includegraphics[width=0.92\textwidth]{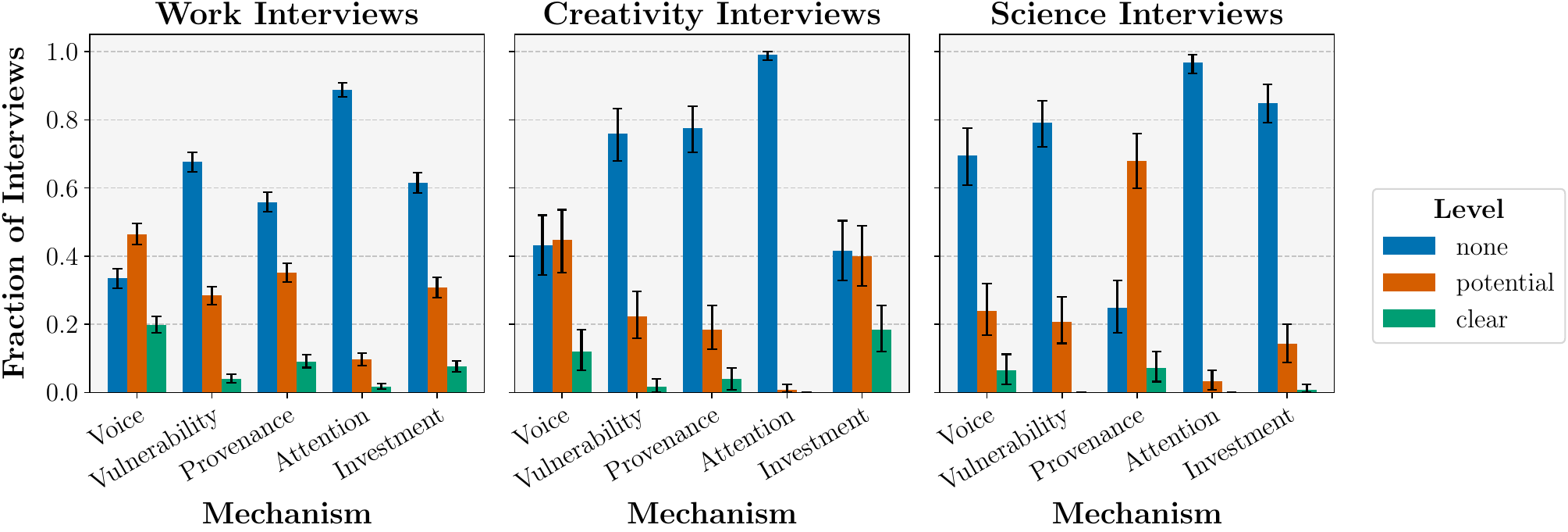}
\caption{Opacity mechanisms by professional group.}
\label{fig:groups}
\end{figure}

Beyond corpus-wide prevalence, the specific mechanisms that professionals discuss vary systematically by domain (Figure~\ref{fig:groups}). Voice opacity was most prevalent among general workforce participants (66.4\%) and creative professionals (56.8\%), and notably lower among scientists (30.4\%; $\chi^2 = 62.52$, $p < .001$, $V = 0.22$). Post-hoc pairwise comparisons confirm that scientists differ significantly from both other groups (vs. workforce, $p < .001$; vs. creatives, $p < .001$), while workforce and creatives do not differ significantly from each other ($p = .13$). The split runs primarily between scientific and non-scientific work.

Provenance opacity showed the inverse pattern. It was highest among scientists (75.2\%), moderate in the workforce (44.2\%), and lowest among creatives (22.4\%; $\chi^2 = 72.08$, $p < .001$, $V = 0.24$, with all three pairwise differences significant after Bonferroni correction, $p < .001$). One scientist captured the logic in their interview: ``I will ask for references and then check them\ldots about half the time the AI will give quotes that don't exist. The possibility of nonsense means I can't send an application in, as it will be reviewed by people who know the subject'' [science\_0121].

Investment opacity was most prevalent among creative professionals (58.4\%), compared to 38.4\% in the workforce and 15.2\% among scientists ($\chi^2 = 49.68$, $p < .001$, $V = 0.20$, with all three pairwise differences significant after Bonferroni correction, $p < .001$). Governance frameworks around credit and compensation in creative AI work remain contested \citep{chi2025creative}.

These group differences likely reflect what each field treats as the cue surface of professional credibility. Scientific writing rewards evidentiary restraint over expressive individuality; creative fields valorize visible craft and effort; workforce communication depends on distinctive style as a signal of engagement. Workers tend to defend the mechanism whose loss would most damage their domain's specific credibility infrastructure.

\subsection{Identity/Labor Asymmetry}

\begin{figure}[htbp]
\centering
\includegraphics[width=0.92\textwidth]{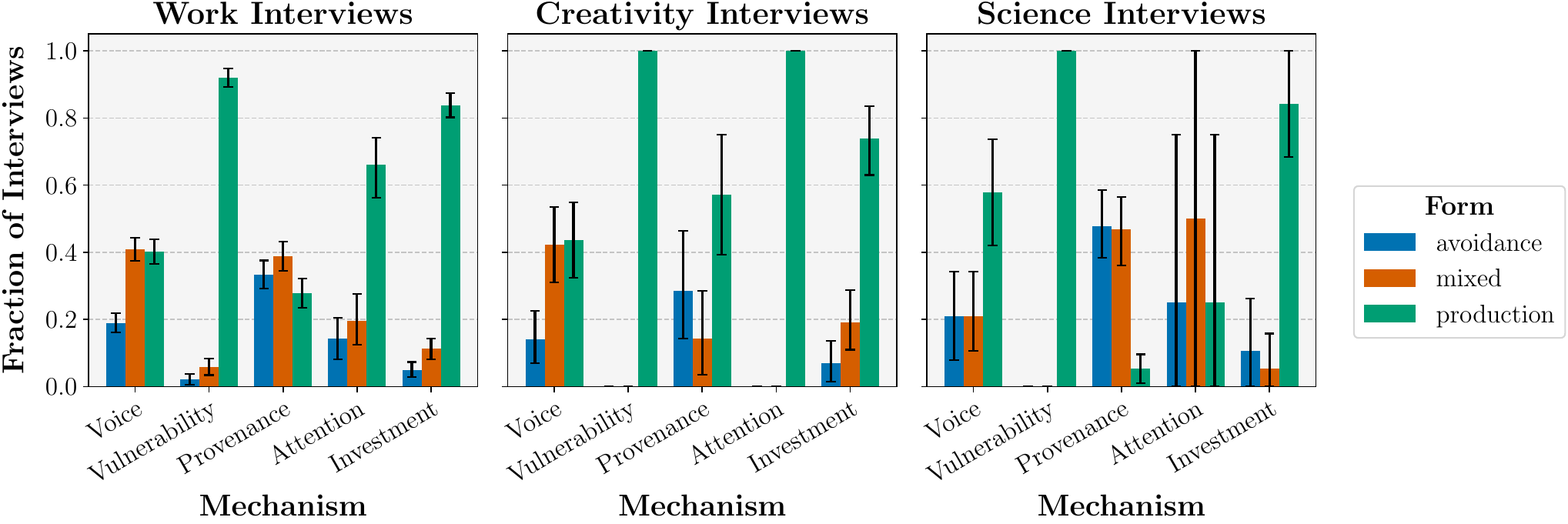}
\caption{Behavioral orientations by mechanism. Bars show fraction of cases reflecting avoidance, mixed, or production orientations.}
\label{fig:form}
\end{figure}

Beyond prevalence, the most striking finding is in workers' behavioral orientation, which varies dramatically by mechanism type. Figure~\ref{fig:form} displays the distribution of production, avoidance, and mixed orientations across mechanisms.

Provenance opacity showed the highest avoidance:production ratio at 1.39:1 (avoidance cases per production case), the only mechanism where avoidance exceeded production overall. This aggregate is driven by scientists (9:1) and workforce (1.2:1); for creatives the ratio inverts (0.5:1, $n = 28$), consistent with provenance's lower role in creative credibility infrastructure. When professionals defend provenance, they describe practices that preserve their capacity to stand behind the work, such as verification, citation-checking, or refusal to sign material they have not finalized.

The labor-relevant mechanisms, however, present a different picture. Vulnerability opacity was near-universally produced: 93.2\% of cases involved creating or allowing opacity, with only 1.8\% avoidance (ratio 0.02:1). Participants described hiding knowledge gaps and presenting façades of competence with little apparent concern. Investment opacity followed the same pattern, with 82.4\% production against 5.5\% avoidance (ratio 0.07:1). Cases such as the software developer who ``had AI do all the work'' while appearing to ``lead'' the project were common and largely unreflective. Attention opacity, though rarer overall, was also predominantly produced (65.0\% production, 14.5\% avoidance, ratio 0.22:1).

The pattern aligns with the identity/labor distinction introduced in \S 3. Identity mechanisms (voice, provenance) showed 464 cases of production against 344 of avoidance (ratio 0.74:1). Labor mechanisms (vulnerability, attention, investment) showed 822 cases of production against only 50 of avoidance (ratio 0.06:1). The difference between these ratios is highly significant ($\chi^2 = 315.01$, $p < .001$, $\phi = 0.43$). Because a single transcript can contribute observations to both mechanism classes, we corroborate this pooled test with the cluster-robust checks described above. A GEE regression yields an odds ratio of 10.7 (95\% CI $[8.2, 13.9]$, $z = 17.58$, $p < .001$), while the transcript-level bootstrap yields a 95\% CI of $[9.2, 16.8]$. The avoidance-to-production ratio is therefore roughly twelve times higher for identity mechanisms than for labor mechanisms. This asymmetry reveals selective stigma management: AI use is discreditable when it threatens authorship and voice, and acceptable (or even desirable) when it conceals the labor behind the performance.

\subsection{Activities and Opacity Mechanisms}

\begin{figure}[t]
\centering
\includegraphics[width=0.9\textwidth]{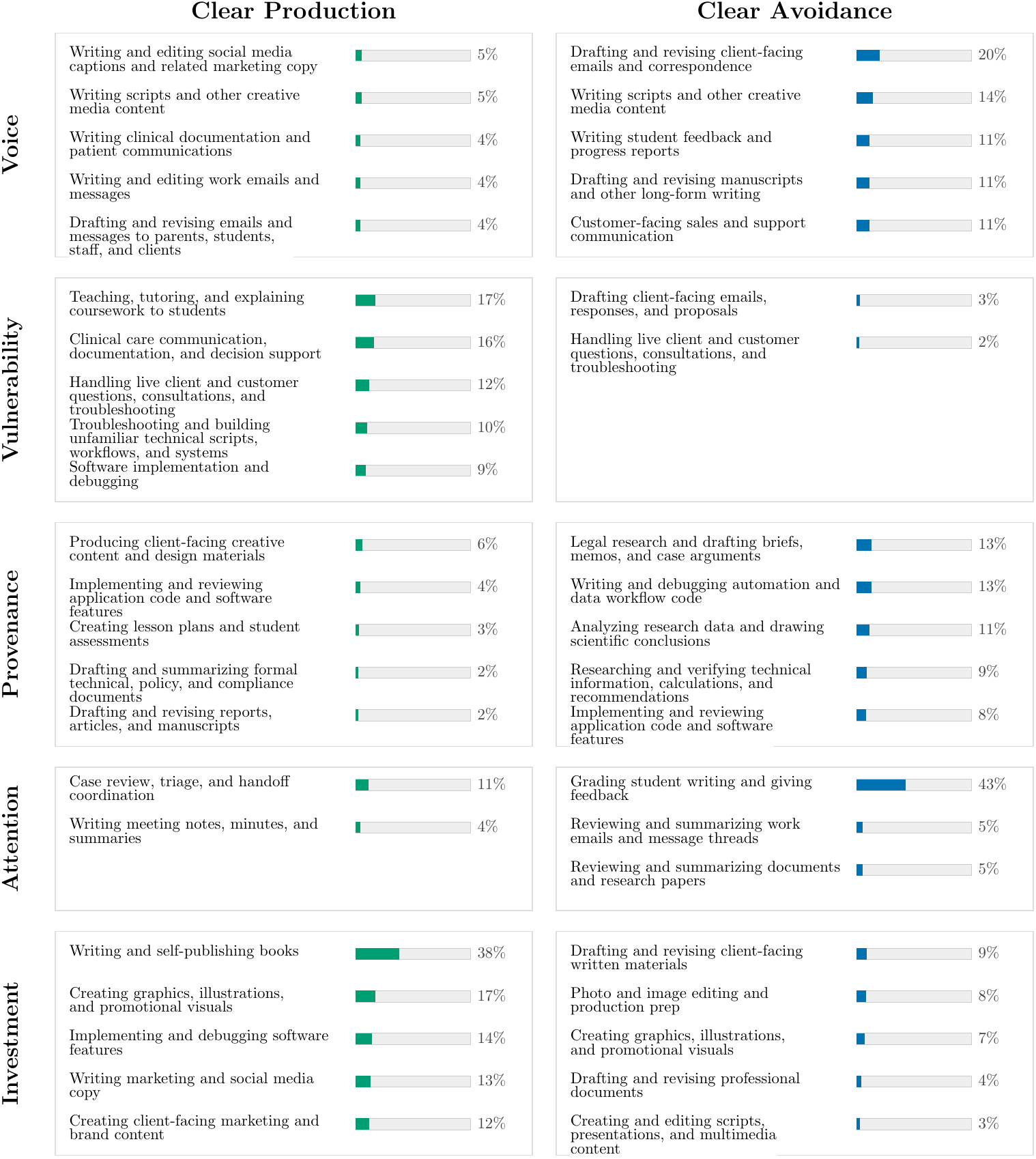}
\caption{Top activity clusters by opacity mechanism and behavioral form.  Bars show the share of each mechanism's clear-production (left) or clear-avoidance (right) cases that fall in the activity cluster.}
\label{fig:activities}
\end{figure}

To identify work contexts, we summarized the task associated with each coded mechanism, embedded the summaries with \texttt{all-mpnet-base-v2}, clustered them by mechanism using $k$-means ($k = 8$ to $18$ per mechanism), and labeled the resulting clusters with GPT-5.4. Figure~\ref{fig:activities} reports the most common activity clusters by mechanism and orientation.

We find that avoidance dominates only in the provenance mechanism, a pattern whose forensic logic we return to in \S 4.4. The close readings that follow turn to production, where the analytic value of our typology become most visible.

The production of \textit{attention} opacity concentrates in case reviews, meeting documentation, and summarizing research. These are contexts where engagement can be delegated without immediate consequences. Attention opacity avoidance, by contrast, is dominated by a single activity, grading student writing (43\%), where teachers refuse to let the appearance of having read substitute for actually reading. The production side shows how AI-generated notes can distort both meeting participation and the record on which later coordination depends. A project lead at a software company describes how AI note-takers reshape meeting participation:

\begin{quote}
``Sometimes, people are less involved because they know they can rely on the AI to give them the meeting notes at the end. The most extreme example is not joining the call at all\ldots\ I do review the overview at the end to ensure those notes are accurate, especially if people rely on them. [The AI is] leaving tasks unassigned when they were assigned, and creating tasks for things that were just notes in the call. Lastly, the AI could miss delegating things to people as well.'' [work\_0681]
\end{quote}

AI note-takers reshape attendance through a feedback loop. Because workers expect the system to capture the meeting, they participate less and, at the extreme, do not join the call at all. Yet the record they rely on cannot reliably preserve the meeting's allocation of responsibility--that is, which tasks were assigned and who agreed to carry them out. The fewer who attend, the fewer who can challenge the AI's account. 

The production of \textit{investment} opacity dominates across creative and professional contexts, such as self-publishing books, creating graphics and illustrations, and writing marketing copy. But the practice also occasionally scrambles the producer/avoider distinction at the heart of our typology. One speaker, transitioning from manual labor into administrative responsibilities, describes editing AI-generated reports before sending them to colleagues:

\begin{quote}
``I'll take what the AI wrote, remove any filler, change any words I wouldn't naturally use in my own speech, and heck, maybe even add in a grammatical error just so that my colleagues know I'm human and we all make mistakes. It's dangerous territory to try and appear perfect all the time because people start expecting certain standards from you, or start to look at themselves and think they're not good enough.'' [work\_0941]
\end{quote}

The inserted grammatical error mentioned by this interviewee is part of a recursive process, a sign-of-being-a-sign-of-effort. Opacity, here, is masked by re-coupling output to counterfeit effort cues: producers manufacture the markers of human effort that AI-generated text otherwise lacks. Yet making AI output pass as average human text turns out to require substantial new effort, including linguistic detection of AI tells and calibrated error-insertion. Investment opacity does not eliminate labor so much as migrate it from the visible product to invisible upkeep. Notably, the participant engages in this process precisely because otherwise ``people start expecting certain standards from you.'' They produce opacity to manage the audience's future interpretive standards, on behalf of a workplace whose norms have not yet formed.

\textit{Provenance} avoidance dominates the mechanism overall and concentrates in high-accountability domains such as legal research, brief drafting, debugging automation code, and analyzing research data, where the producer's name carries subsequent consequence and delegation is risky. The rare cases of provenance \textit{production}, however, reveal a structural consequence that the avoidance accounts do not. The most telling case in the corpus comes from an interviewee who is detecting it, rather than producing it. The owner of a small handyman business describes what happened when they realized their graphic designer was laundering AI output through her professional identity:

\begin{quote}
``Since I realized that our graphic designer was just using AI, I've realized that she is not choosing to think on her own about projects very much at all. I have to imagine that I'm actually prompting the AI myself, so I have to be very clear and specific with what I ask from her, as I imagine she is just copying/pasting it into some sort of AI. Really ought to just skip the middleman, to tell the truth.'' [work\_0763]
\end{quote}

Most provenance discourse is about workers managing their own AI use; this account is from an audience that has seen through such management. The detection does not trigger confrontation. The owner does not fire the designer, or demand disclosure; instead, they silently re-classify her in their own working model, moving her from a worker with judgment to \textit{middleman}, a passthrough between their instructions and the model's output. The reorganization is unilateral and undisclosed, explained as having to ``imagine that I'm actually prompting the AI myself.'' The front and back stage do not collapse in any visible way here. What collapses is the designer's epistemic standing in the owner's mental model of the workflow, with no occasion for repair. The closing sentence shows how provenance opacity, once detected, becomes a precursor to labor disintermediation: the worker, treated as a passthrough to AI, is silently routed around inside the workflow before any formal decision to cut her. This rerouting will likely only be visible to the designer once her position has already eroded. The case suggests how, aggregated across many such silent decisions, workplaces may begin to sort workers from the labor inputs they once embodied.

\textit{Vulnerability} opacity production appears across contexts where real-time competence displays matter, including teaching and tutoring (17\%), clinical care communication (16\%), and technical troubleshooting (10\%). It also concentrates in the early career, where the audience whose uncertainty-detection capacity is being managed is, precisely, the senior expected to detect and correct it. A medical student describes their use of OpenEvidence in exactly these terms:

\begin{quote}
``I use OpenEvidence and ask it questions as if I'm talking to a senior resident or attending and asking them questions I'm too nervous to ask in real life\ldots sometimes I use it to check over my differential or assessment and plan before presenting to my attending and it makes me feel more confident that what I'm saying makes medical sense.'' [work\_0264]
\end{quote}

The participant is replacing a person whose institutional role is to receive novice uncertainty and convert it into instruction, rather than replacing reference materials. The AI substitute is engineered to bypass the relational stakes, including embarrassment, the appearance of unpreparedness, and the small ritual humiliations through which clinical apprenticeship has historically proceeded. By using AI to ``check over'' their assessment-and-plan before presenting to the attending, they replace what was formerly a back stage encounter with a senior with a real-time pocket resource. The senior, no longer a teammate who would absorb novice questions, becomes another front stage audience they must simply perform competence to.

Vulnerability opacity is therefore a form of \textit{face-work} under conditions of automated rehearsal, in which the novice preserves the competent front before seniors by moving the risky moment of not-knowing into a private interaction with an LLM. Vulnerability is managed through a substitute that simulates the senior's epistemic role while stripping out the senior's recognition function. Vulnerability opacity, thus, dissolves the social occasion through which the gap would be named and addressed.

Finally, \textit{voice} avoidance concentrates in client-facing correspondence (20\%) and creative writing (14\%)---contexts where distinctive style indexes the person behind the interaction. The most theoretically revealing cases, however, are the productions routed through avoidance. An Early Years Educator, describing how they now write daily learning observations for parents, explains:

\begin{quote}
``I have fed an AI examples of my own personal observations as an example template, and the AI now uses this template to write my observations for me\ldots\ I like it to use my style of writing so that my work doesn't \textit{look} AI generated, there's a bit of stigma around it, I don't tell my colleagues that I use it\ldots\ I do worry that\ldots\ if I write my own, someone will notice the difference in quality! Which is a little silly, I did use my own language and observations as the template after all.'' [work\_0245]
\end{quote}

The participant has constructed a reflexive loop. They curate a corpus of their own past front-stage performances, hand it to the model as a stylistic template, and then animate the model's outputs as their own. They remain the principal and the animator, but authorship has been displaced onto a statistical compression of their earlier self. We might call this \textit{autoventriloquism}, a form of voice opacity produced through self-impersonation with the model functioning as a stylistic mirror conditioned on prior performances of the self. Personhood cues like ``my style of writing'' become signs of an earlier self, reanimated in the present. Voice becomes an archived front, rather than a present expressive capacity. The audience can no longer read voice as evidence of present engagement, while the worker, having set the archived double as the standard, finds their unmediated writing falling short of it in real time.

\subsection{From Asymmetry to Inspectability}

While the identity/labor asymmetry tells us which mechanisms professionals defend, it does not tell us why. Two structural features of our corpus may provide further explanation. AI workplace opacity is more accurately understood as a new social organization of \textit{inspectability}, in which workers defend the signs that can be detected while routing disclosure differently across the audiences who hold them to account.
 
The first feature is a forensic asymmetry. Identity mechanisms leave detectable residue such as em-dashes and the verdicts of AI-detection software, while labor mechanisms have no comparable cue surface. There is no detector for effort that was not expended, and no software that scans a finished memo for the hesitations it does not contain. A technical writer who serves multiple regulated clients articulates the resulting economy:

\begin{quote}
``They all want me to use AI, but ensure the writing cannot be detected by AI-detecting software. It is a paradoxical situation\ldots\ I tell the AI to take each citation and tell me the 1st sentence of the first body paragraph. So far, that always works.'' [work\_0240]
\end{quote}

The ``paradoxical situation'' resolves once one disaggregates audiences. Clients want prose that will not register as machine-written to the third-party detector they fear, rather than human-written prose as such. The writer therefore performs two laundering operations. Outwardly, they scrub the cue surface so the artifact passes the client's detector. Inwardly, they run their own forensic test on the AI to catch hallucinations before the client does. Yet no comparable apparatus is contemplated for the labor the AI is doing on the article body, where hours of research and synthesis are accepted without question because that labor leaves no visible trace.

The same logic recurs across the corpus. An Early Years Educator deletes ``the MANY MANY em-dashes'' from their AI-generated observations because they are ``super obvious'' giveaways [work\_0245]. A developer who generated polished business descriptions reports that ``nobody seems to question the source'' [work\_0278]. A worker who ``had AI do all the work'' on a database design notes that colleagues now come to them ``for domain knowledge'' while they ``did that quite leisurely'' [work\_0693]. Where detection is possible, opacity is policed. Where it is not, opacity is produced.
 
The second feature is that this forensic substrate is itself unevenly distributed across audiences. Our cases show why workers conceal selectively---hiding AI use from managers and evaluators while remaining open with peers: peers often share the same detection capacities the worker possesses, while audiences above lack them. The audience-relative pattern thus has a forensic substrate, not only a relational one. A software developer makes the segmentation explicit:

\begin{quote}
``The devs openly share features and tooling with each other. We completely hide it from management and our product people\ldots\ I didn't tell my employer that the 16 minute spike took 16 minutes. I kept the day to myself to give myself an extra long weekend.'' [work\_0747]
\end{quote}

The peer collective functions as a \textit{performance team} in Goffman's sense \citep{goffman1959presentation}, a region where fellow practitioners actively coordinate the very practice they jointly conceal from above. Mutual disclosure in the back stage enables coordinated nondisclosure on the front stage. The producer/avoider distinction collapses here. Coded in the aggregate as a producer, work\_0747 is laterally an avoider: they openly share with peers the AI use they hide from management. Orientation, in other words, is a property of the worker-audience dyad rather than of the worker alone. The same juniors who hide vulnerability from their attendings admit it ``alone with some residents or students'' [work\_0472], and the same consultants who conceal time savings from fixed-price clients share their workflow with peers [work\_0096]. What looks in the aggregate like selective stigma management is, at the relational level, coordinated audience segregation in which different audiences are entitled to different disclosures.

\subsection{The Logic of Accounts}

LLM-assisted coding captures prevalence and behavioral orientation, but accounts also reveal \textit{practical reasoning}, the way professionals justify their practices \citep{scott1968accounts}. As described above, we hand-coded 207 cases for accountability- and efficiency-oriented justifications, which are not mutually exclusive.
 
\begin{figure}[htbp]
\centering
\includegraphics[width=\textwidth]{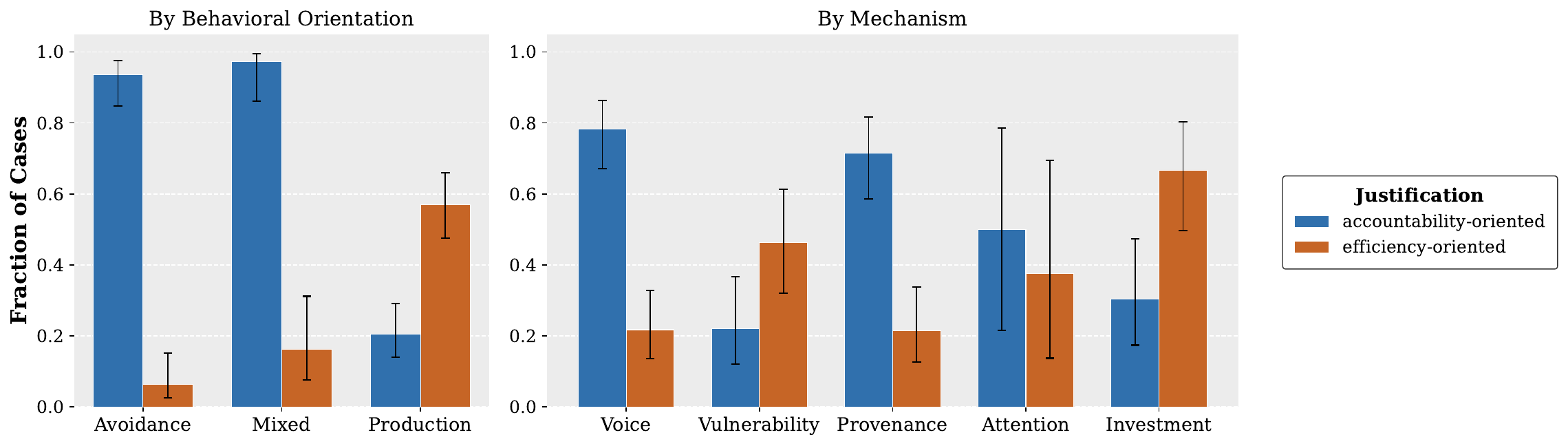}
\caption{Justification types across hand-coded cases ($n = 207$). Bars show the fraction of cases containing accountability-oriented and efficiency-oriented justifications. Categories are not mutually exclusive. Error bars are 95\% Wilson confidence intervals.}
\label{fig:justifications}
\end{figure}

Accountability justifications appeared in 56.5\% of cases, efficiency in 34.3\%. Justification type varied sharply by behavioral orientation and by mechanism (Figure~\ref{fig:justifications}). 93.7\% of avoidance cases contained accountability justifications, compared to only 20.6\% of production cases. Accountability justifications appeared in 78.3\% of voice cases and 71.4\% of provenance cases, but in only 22.0\% of vulnerability cases and 30.3\% of investment cases.

The 20.6\% of production cases that do invoke accountability tend to cluster where workers can claim the labor as their own expertise. A freelance editor shows the move:

\begin{quote}
``If I feel like that is how the text should read, I will use AI's suggestion. I prefer to blend the work as I feel like I'm cheating a bit otherwise. The author is paying me to edit their work, so using AI sometimes feels like the easy route. I know that the author wouldn't be able to edit their own work, even using AI, as it is a skill.'' [work\_0361]
\end{quote}

The editor invokes professional obligation (``the author is paying me''), acknowledges efficiency gains (``the easy route''), and resolves the tension by reasserting that editing ``is a skill.'' Accountability and efficiency justifications coexist when the worker can claim the labor as their own expertise.
 
The general finding behind this asymmetry is more consequential than any single case. Resisting AI use in efficiency-oriented workplaces is untoward and must be accounted for through responsibility, liability, or trust, while producing labor opacity passes without comment. In these accounts, efficiency operates as the unmarked moral default; what requires explanation is often the refusal to obscure effort, not the obscuring itself.
 
This reflexivity asymmetry recurs at the salience level. Workers articulate opacity most explicitly for voice (29\% of voice-present cases are coded as ``clear'' rather than ``potential'') and least for vulnerability (11\%) and attention (14\%). When AI fills in for uncertainty or attendance, workers describe it as workflow efficiency---``I turn to AI when I'm stuck,'' ``I let AI handle the meeting notes''---without naming the social work being done. Reflexivity, in other words, tracks vocabulary. Workers recognize concealing AI's style as something that calls for justification, so they justify it. The practices the workplace has no language for, such as covering uncertainty, workers enact just as readily, and seemingly without recognizing them as opacity.

\section{Discussion and Conclusion}

This article has shown how GenAI reorganizes the cue surfaces through which workers become accountable to one another. The most consequential finding is that workers did not treat AI mediation as uniformly problematic. They were more likely to protect identity-bearing cues, especially voice and provenance, while allowing labor-bearing cues such as effort, attention, and uncertainty to disappear into otherwise acceptable outputs. Voice and provenance remained contestable because they attach an artifact to a recognizable source: a colleague can notice that a message no longer sounds like its sender, or a client can question whether a designer produced an image. Effort, attention, and uncertainty, by contrast, receded into the completed task and became accountable mainly when something went wrong.

This asymmetry reflects the output-centered organization of contemporary work. Workplaces commonly treat the deliverable as evidence that work occurred: emails sent, tickets closed, tasks completed, and so on. These artifacts stand in for the process that produced them. Output-centered workplaces, in other words, rely on a practical inference: if the deliverable is adequate, the work behind it has been adequately performed. GenAI weakens that inference by making the same deliverable compatible with very different levels of human effort, attention, uncertainty, and judgment.

The accounts show the normative consequence of this weakened inference. When AI changed backstage labor inputs while preserving the deliverable, workers could describe the change as efficiency rather than concealment. The artifact was created, the task was completed--and the production process remained secondary. By contrast, workers often limited AI use when voice or provenance was at stake, even when the resulting output was technically adequate. They cared whether the communication still sounded like them and whether they could claim and defend the work as their own. The boundary of acceptable AI use in these cases was therefore not automation itself, but whether automation disrupted the relation between output and the accountable subject who produced it.

The consequences for policy are, in sum, less about policing AI use than about defining required forms of human participation. Generic disclosure policies, focused only on whether AI was used, do little to prevent the output-centered infrastructure that makes labor opacity easy to produce in the first place. GenAI makes visible a workplace obligation that output-centered systems often leave unstated: workers are expected not only to deliver an acceptable result, but also to participate in producing it in specific, recognizably human ways. The harder governance problem is to specify which acts are non-delegable, which forms of human involvement must remain inspectable despite an adequate output, and which audiences are entitled to know the difference.

Finally, it is worth noting our corpus's own structure illustrates our argument. Participants itemized to an AI interviewer the concealment they withhold from managers and clients. Claude acted as an interlocutor with no stake in their competence, no sanctioning capacity, and as such constituted an audience entitled to admissions that workplace audiences are not. That the interviews elicited concealment practices at all is itself evidence of audience-calibrated disclosure.

Our study has three main limitations. First, the corpus is recruited, self-reported, and AI-mediated. Disclosure to Claude is not equivalent to disclosure to a colleague, and the transcripts capture workers' accounts rather than observed workplace interactions or audience responses. Second, our subsamples are unbalanced, so domain comparisons should be read as broad contrasts rather than fine-grained occupational claims. Third, primary coding was LLM-assisted; blinded human coding of a 50-transcript subsample (\S 4) shows substantial human--model agreement, but the subsample is small. We release the prompts, labels, validation materials, and code to support further validation and extension.

\section*{Data and Code Availability}

The Anthropic Interviewer transcripts analyzed in this paper are publicly available under CC-BY at \url{https://huggingface.co/datasets/Anthropic/AnthropicInterviewer}. All analysis code, LLM coding prompts, the hand-coded justification sheet, the validation coding protocol with its sampling and agreement scripts, and the figure-generation scripts used to produce the results reported here are released at \url{https://github.com/dlab-projects/fabricated_front}. The exact coding prompt is released verbatim at \url{https://github.com/dlab-projects/fabricated_front/blob/main/prompts/opacity_typology_coding_v3.txt}.



\end{document}